# Migration of phthalate plasticisers in heritage objects made of poly(vinyl chloride): mechanical and environmental aspects

Sonia Bujok[a*], Tomasz Pańczyk[a], Kosma Szutkowski[b], Dominika Anioł[b], Sergii Antropov[a], Krzysztof Kruczała[c], Łukasz Bratasz[a]

[a] Jerzy Haber Institute of Catalysis and Surface Chemistry Polish Academy of Sciences, Niezapominajek 8, 30-239 Kraków, Poland

[b] NanoBioMedical Centre, Adam Mickiewicz University in Poznań, Wszechnicy Piastowskiej 3, 61-614 Poznań, Poland

[c] Faculty of Chemistry, Jagiellonian University, Gronostajowa 2, 30 - 387 Kraków, Poland

***Corresponding author:** Sonia Bujok
e-mail: sonia.bujok@ikifp.edu.pl, phone: +48 12 6395 190, postal address: Jerzy Haber Institute of Catalysis and Surface Chemistry Polish Academy of Sciences, Niezapominajek 8, 30-239 Kraków, Poland

## Abstract
To clean or not to clean? The solution to this dilemma is related to understanding the plasticiser migration which has a few practical implications for the state of museum artefacts made of plasticised poly(vinyl chloride) - PVC and objects stored in their vicinity. The consequences of this process encompass aesthetic changes due to the presence of exudates and dust deposition, an increase in air pollution and the development of mechanical stresses. Therefore, this paper discusses the plasticiser migration in PVC to provide evidence and support the development of recommendations and guidelines for conservators, collection managers and heritage scientists. Particularly, the investigation is focused on the migration of the ortho-phthalates representing the group of the most abundant plasticisers in PVC collections. The predominance of inner diffusion or surface emission (evaporation) determining the rate-limiting step of the overall migration process is considered a fundament for understanding the potential environmental and mechanical risk. According to this concept, general correlations for various ortho-phthalates are proposed depending on their molar mass with the support of molecular dynamics simulations and NMR diffusometry. The study reveals that for the majority of the PVC objects in collections, the risk of accelerated migration upon mild removal of surface plasticiser exudate is low. Thus, surface cleaning would allow for diminishing dust deposition and air pollution by phthalate-emitting objects in a museum environment. Bearing in mind simplicity and the need for fast decision-supporting solutions, the step-by-step protocol for non-destructive identification and quantification of plasticisers in objects made of or containing plasticised PVC, determination of the physical state of investigated artefacts and rate-limiting process of plasticiser migration is proposed.

## Highlights
- Plasticiser migration-induced mechanical and environmental risk to museum artefacts
- Molecular dynamics for study of plasticiser diffusion in poly(vinyl chloride)
- Nuclear magnetic resonance diffusometry for solid-state diffusivity studies
- Attempt of generalisation of ortho-phthalates migration in poly(vinyl chloride)
- Evidence-based hints for conservators and decision-makers in heritage institutions

**Keywords:** heritage science, phthalate plasticisers, poly(vinyl chloride), molecular dynamics, NMR diffusometry

## 1. Introduction

Preservation of heritage objects made of plasticised polymers is a challenge for modern and contemporary art conservation practices. One of the most problematic polymers is a poly(vinyl chloride) – PVC. The main issue is the lack of stability in terms of plasticiser migration leading to visual changes, e.g. the presence of surface exudates leading to stickiness of the surface and a higher rate of dust deposition (King et al., 2020). These changes contribute to the loss of aesthetic value but also cause deformations and embrittlement (King et al., 2020; Rijavec et al., 2020; Shashoua, 2001, 2003). While some investigations are focused on the consequences of plasticiser migration related to surface cleaning, the physical (mechanical) instability of the objects is not widely discussed (Apchain et al., 2022; King et al., 2020; Royaux et al., 2017). Generally, plasticiser migration is considered a two-step mass transport process of bulk diffusion and subsequent surface emission (evaporation) of plasticiser molecules or phase separation (King et al., 2020; Wei et al., 2019). The diffusion-evaporation model describing the migration of small molecules in a polymer matrix is commonly applied for industrial modelling (Ekelund et al., 2010; Linde and Gedde, 2014; Wei et al., 2019) and was mentioned in the heritage science field as the dominant process of deterioration of objects made of plasticised polymers (Apchain et al., 2022; King et al., 2020; Royaux et al., 2017; Shashoua, 2001, 2003). Depending on the rate of each step, the overall plasticiser loss can be diffusion- or evaporation-controlled, which is described by the dimensionless $R=FL/D$ parameter that defines the relation between diffusion coefficient $D$, surface emission coefficient $F$, and the object's thickness $L$ (Wei et al., 2019).

The predominance of plasticiser diffusion or evaporation has practical implications on the presence of surface exudates, i.e. if the plasticiser evaporation is the rate-limiting step ($R \ll 1$) it may lead to the accumulation of the molecules on the object surface and the creation of the thin film, which may not be macroscopically visible and could rather result in a tacky surface (King et al., 2020; Wei et al., 2019). In such cases, there is uniform plasticiser distribution in the bulk of the material. On the contrary, diffusion-controlled migration ($R \gg 1$) results in a gradient of a plasticiser concentration making the surface layer of the object more rigid, brittle and thus less resistant to rapid external loading, i.e. high impact during transportation (Wei et al., 2019). Although it was not decisively shown whether the tackiness of the surface depends on the relative rate of plasticiser evaporation and diffusion ($R$ parameter), or just on the rate of evaporation ($F$), it is clear that the determination of the rate-limiting step of the plasticiser loss is crucial for the qualitative assessment of mechanical damage risk to heritage objects. It is also relevant for the discussion of whether surface cleaning affects the rate of plasticiser migration. So far, most researchers believe that the process is not affected by plasticiser removal when the migration is evaporation-controlled. On the other hand, there is a conviction among some researchers that if the diffusion is a slower step, surface cleaning of plasticiser traces accelerates the overall plasticiser migration and deterioration rate of the objects due to a larger concentration gradient (King et al., 2020; Shashoua, 2008, 2001).

A question frequently raised by the conservation community is how the removal of the surface exudates affects the deterioration rate of the object, but also if the plasticisers' vapours interfere with other objects stored in the same enclosure and influence their appearance and deterioration rate (Lattuati-Derieux et al., 2013). Studies on the organic air pollutants in the museum collections indicated the presence of phthalate plasticisers and/or their degradation products (Lattuati-Derieux et al., 2013; Mitchell, 2014; Schieweck, 2020), which were also detected in the particulate matter or settled dust in an indoor environment (Fujii et al., 2003; Kang et al., 2021; Song et al., 2015). Considering the environmental aspects of the plasticisers' release, it is worth mentioning that safety aspects are increasingly more relevant in collection management as phthalate plasticisers, especially bis(2-ethylhexyl) phthalate – DEHP, were classified as hazardous.

Undoubtedly, more experimental and theoretical research is needed to develop evidence-based recommendations for surface cleaning, storing objects and air ventilation, particularly including phthalate-emitting objects, that is relevant for risk assessment to the overall collection as well as air quality management in memory institutions.

To identify the rate-limiting step of the plasticiser loss in heritage PVC objects, accelerated ageing experiments (70-80 ºC) were performed and investigated by gravimetric changes combined with spectroscopic, chromatographic, thermogravimetric, or microscopic techniques (Apchain et al., 2022; Royaux et al., 2017; Shashoua, 2003). Despite the long-term ageing procedures, their extrapolation to the typical display/storage (20 ºC) or cold storage conditions (10 ºC) in memory institutions remains a tentative issue. Moreover, proposing universal conservation guidelines for the entire PVC heritage collection is challenging due to the diversity of the plasticisers found in PVC-based objects (King et al., 2020; Rijavec et al., 2022, 2020; Saad et al., 2024). Among various plasticisers used to soften PVC, DEHP, bis(2-ethylhexyl) terephthalate - DOTP, diisononyl phthalate - DINP, and di-*n*-butyl phthalate - DBP are the most frequently determined in PVC collections originating from the 1960s-1970s (Gassmann et al., 2023; Jadzińska, 2019, 2017; Klisińska-Kopacz et al., 2019; Rijavec et al., 2020; Royaux et al., 2017). Based on the available literature, migration of DEHP and DINP is assigned to be rather an evaporation-controlled process (Apchain et al., 2022; Royaux et al., 2017). However, the rate-determining step for smaller ortho-phthalate, e.g. DBP, has not been widely investigated so far despite its presence in PVC heritage objects identified as unstable with multiple signs of damage, e.g. mannequins created in the 1970s by Tadeusz Kantor for series of *The Dead Class* plays in the Cricot 2 Theatre (Figure 1) (Jadzińska, 2019, 2017; Klisińska-Kopacz et al., 2019). Recent studies also reported the presence of other lower molar mass ortho-phthalates, e.g. diethyl phthalate – DEP, diisobutyl phthalate – DiBP or dihexyl phthalate – DHP, which constitute about 10% of the PVC heritage collections considered as highly unstable (Gassmann et al., 2023; Rijavec et al., 2022).

In response to the demand for fast and simple plasticiser determination, tools based on machine-learning-assisted spectroscopic analyses were developed for museum practitioners to support their decision-making process concerning conservation treatment (Rijavec et al., 2022; Saad et al., 2024). In combination with the vast study on diffusion-evaporation migration of phthalate plasticisers, evidence-based conservation practices and preventive actions will be better supported.

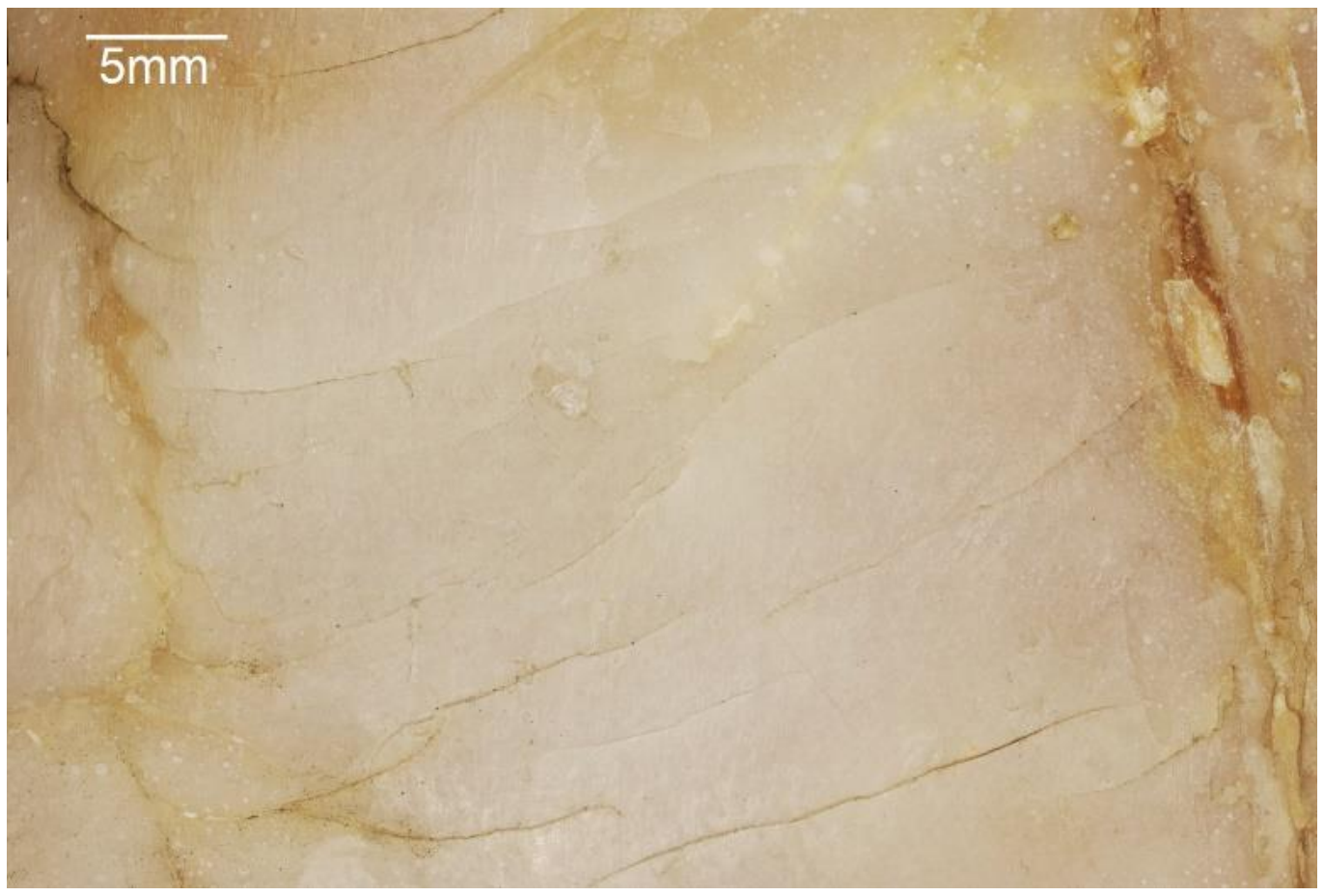

**Figure 1.** High-resolution macroscopic scan of the PVC-DBP mannequin surface with cracks and scratches (Tadeusz Kantor, Right leg of the girl in the black dress from *The Dead Class (Umarła Klasa)*, late 1980s, collection of the Cricoteka – Tadeusz Kantor Art Documentation Centre, Cracow, Poland).

Despite numerous studies on the diffusion of phthalate plasticisers in PVC and a variety of methods for the determination of their diffusion coefficients *D*, significant discrepancies between obtained *D* values are observed depending on the methodological approach (Chen et al., 2022; Tüzüm Demir and Ulutan, 2013). The most commonly used method for determining *D* is based on the extraction of the plasticiser into a selected organic solvent, which potentially overestimates obtained *D* values due to the simultaneous swelling of the material that enhances further diffusion (Brouillet and Fugit, 2009; Sarath Josh et al., 2012). Therefore, a few computational attempts based on molecular dynamics (MD) simulations were made to estimate plasticisers' or other small molecules' diffusion in the polymer matrices (Bharadwaj and Boyd, 1999; Han and Boyd, 1994; Iscen et al., 2023; Mao et al., 2023; Müller-Plathe, 1991; Pan et al., 2019; Pant and Boyd, 1993, 1992). Due to the specific arrangement of the plasticised PVC on the molecular level, i.e. plasticiser molecules are trapped in the cavities formed by the polymer chains, three diffusion regimes were distinguished depending on the time scale. Local mobility of plasticiser molecules within the cavities is considered short-time diffusion (sometimes referred to as self-diffusion), which is followed by the anomalous diffusive regime, where *D* is time-dependent. Anomalous diffusion is related to the in-cavity motion of plasticiser affected by the dynamics of the polymer chains (special inhomogeneity of the cavities distribution). Diffusivity in both these regimes does not follow Einstein's relation (1)

$$\langle |r(t) - r(0)|^2 \rangle = MSD = 6Dt \quad (1)$$

where *r(t)* is particle position at a given time with reference to initial particle position *r(0)* and *MSD* refers to mean-square displacement in time *t* (with an exponent of 1 for normal diffusion; for short-time and anomalous *t* exponent is smaller than 1) (Gusev et al., 1994). Long-term diffusive motion (normal diffusion based on the jump of plasticiser molecules between cavities) can be solely reached for simulation times usually above 100 ns, for which Einstein's formula is valid and *D* can be calculated. However, such MD simulations require long computing time, thus, the time scale of hundreds of ns is practically inaccessible to MD, which is the main limitation of this method (Gusev et al., 1994; Mao et al., 2023; Pan et al., 2019). Due to these computational obstacles, a method for the prediction of long-term *D* values based on short-time simulations by fitting the sigmoid function was proposed (Blanco et al., 2018). It should be noted that the simulation results require validation. For this purpose, pulsed-field gradient spin-echo nuclear magnetic resonance (PGSE NMR) spectroscopy is applicable, since this method allows characterisation of the molecular motion in the time range from milliseconds (ms) to seconds. This corresponds to normal (long-term) diffusion (Callaghan et al., 1998; Stejskal and Tanner, 1965).

This paper aims to provide new experimental and theoretical (computational) data and in combination with data reported in industry-oriented literature, describe the plasticiser migration in PVC collections and to transfer this knowledge to heritage scientists and museum practitioners to support the decision-making process and propose evidence-based recommendations, which would be in line with conservation mission and potentially would allow decrease of the environmental risk of phthalate plasticiser emission. In particular, the step-by-step protocol was proposed starting from (I) the non-destructive spectroscopic identification of plasticiser type and content, (II) establishing the state of heritage PVC objects, and (III) the determination of rate-limiting step of specific ortho-phthalate migration. Additionally, the MD simulations and PGSE NMR spectroscopy were evaluated in terms of the accuracy of diffusion coefficient prediction as faster alternatives to typical extraction and accelerated ageing (gravimetric/chromatographic) methods.

## 2. Materials and Methods

Reference plasticised PVC samples were prepared by solvent casting method using powdered rigid PVC (POLANVIL S, Anwil SA, Poland) and di-*n*-butyl phthalate (DBP) plasticiser (>99.5%, Merck KGaA,

Germany). The PVC films of ca. 0.2 mm thickness containing 20, and 30 wt% of DBP as well as unplasticised PVC film were cast from tetrahydrofuran (pure for analysis, Avantor Performance Materials Poland S.A., formerly POCH S.A.) and left under a fume hood to remove residual solvent for at least 2 weeks.

Dynamic mechanical thermal analysis of the PVC samples (ca. 0.2x8x20 $mm^3$) was performed in the tension mode on the Discovery DMA850 (TA Instruments, USA) in the 10-100 °C temperature range with a heating rate of 1 °C/min at 1 Hz frequency under an inert atmosphere. Glass transition temperature $T_g$ was determined as the maximum of the *tan*δ peak (obtained DMTA curves are available in section A1 in Supplementary Materials).

Molecular Dynamics simulations were carried out to gain insights into the molecular structure of the material and determine the diffusion coefficient *D* of the plasticiser. Based on general knowledge about PVC, a fully amorphous structure was assumed. The AMBER force field (Case et al., 2023) was employed for both the PVC and DBP molecules, while partial atomic charges were determined using the RESP procedure (Dupradeau et al., 2010; Vanquelef et al., 2011). The determined density of the model system (30 wt%) at 25 °C (1.296 $g/cm^3$) was in line with the density of PVC containing 30 wt% of DBP reported in the literature (1.291 $g/cm^3$) (Dlubek et al., 2003), which indicates that the material structure of the model system was well described. The simulations were performed using the GROMACS software (Abraham et al., 2015). A comprehensive step-by-step description of the simulation box construction, computational procedures, and settings is available in the Supplementary Materials (section A2).

Nuclear magnetic resonance (NMR) experiments were conducted on an Agilent DD2 14 T spectrometer (Santa Clara, California, USA), equipped with a Doty Scientific DSI-1372 multinuclear probe-head, which has a maximum magnetic field gradient of 30 T/m. The samples were placed in the standard 5 mm NMR sample tubes. The bipolar pulse pair stimulated echo (Dbppste) (Wu et al., 1995) was employed to acquire spin echoes. The diffusion time (Δ) was set to 150 milliseconds, the magnetic field gradient duration (δ) was set to 2 ms, and the magnetic field gradient (*g*) was iterated between 0 and 22 T/m in 32 steps. The data was analysed using MestReNova 14.3, and the diffusion coefficients were fitted using OriginPro 2024 (see section A3 in Supplementary Materials for further details). The experiments were conducted at temperatures ranging from 45 °C to 110 °C.

## 3. Results and Discussion

### 3.1. State of the heritage PVC objects

The diffusivity of the plasticiser in the polymer matrix is affected by factors such as temperature and material composition. Depending on those parameters polymer material can be in the glassy (below $T_g$) or rubbery state (above $T_g$). Therefore, the first crucial aspect of the decision-making process is establishing the state – rubbery or glassy – of PVC objects in collections. To support collection managers and conservators, the correlation of the plasticiser content with $T_g$ was presented in Figure 2 for two types of plasticisers representing heavier (DEHP) and lighter (DBP) ortho-phthalates. In combination with non-destructive spectroscopic analysis of plasticiser type and concentrations (Rijavec et al., 2022; Saad et al., 2024), functions shown in Figure 2 enable simple estimation of the state of the PVC objects.

Firstly, a function correlating $T_g$ with the weight content of DBP was proposed (Figure 2, green linear function fit: $T_g = -1.97C_{DBP} + 358.11$) based on $T_g$ values obtained from DMTA curves (section A1 in Supplementary Materials) and literature data reported by Huang and co-workers (Huang et al., 1993). For comparison, the same correlation for the most common plasticiser in heritage collections – DEHP – was plotted in Figure 2 (purple linear fit: $T_g = -2.40C_{DEHP} + 349.14$) based on literature data (Rijavec et al., 2020).

Statistical examination of historical PVC objects reported by Rijavec and co-workers revealed an average concentration of DEHP and DBP to be ca. 18 wt% and 5 wt%, respectively (Rijavec et al., 2022). As marked in Figure 2 by vertical dot lines corresponding to these values, $T_g$ is ca. 32 °C (305 K) for PVC containing 18 wt% of DEHP and ca. 77 °C (350 K) for PVC with 5 wt% of DBP, which indicates that in both cases representative PVC objects are in the glassy state at the typical display and consequently at lower temperatures in cool storage conditions.

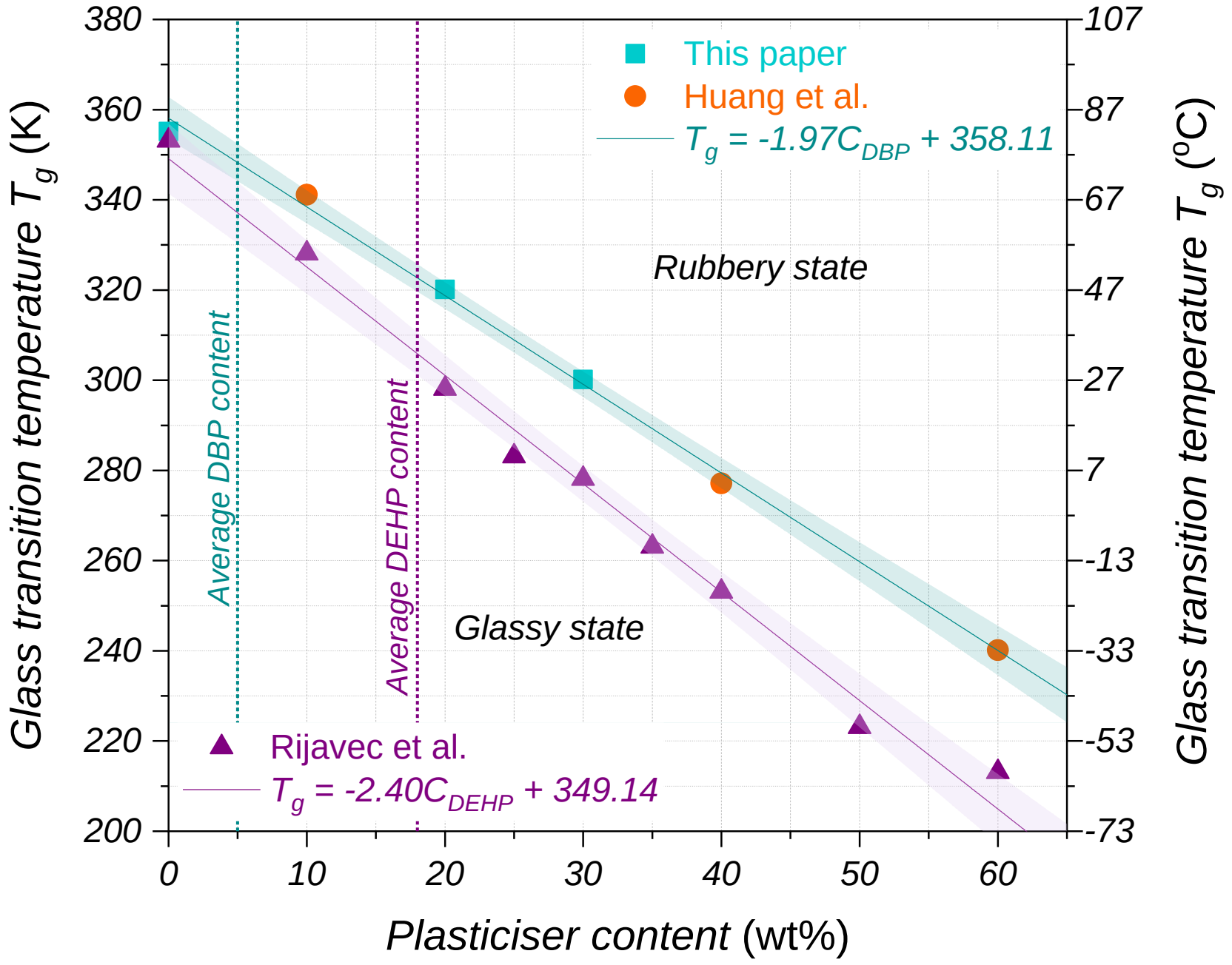

**Figure 2.** Dependence of the glass transition temperature $T_g$ on the weight content of DBP (green linear function fit) and DEHP (purple linear function fit) for PVC. Light green and light purple areas along the fitted functions indicate a confidence band of 95%.

### 3.2. Molecular Dynamics simulations of DBP motion in PVC matrix

As previously discussed (3.1), plasticiser migration is significantly influenced by the physical state of the PVC object, namely whether it is in a glassy or rubbery state. Thus, studies on the diffusion performed at relatively low temperatures, close to the room conditions are valuable from the practical perspective but uncommon in the literature. Although experimental data describing the DBP diffusivity in the glassy state were previously reported by Yuan and co-workers (Yuan et al., 2020), literature data were scarce for highly plasticised PVC with DBP available for methods based on extraction (Brouillet and Fugit, 2009) or β decay (emission) of $^{14}$C-labelled DBP (Griffiths et al., 1984). Based on Figure 2, the minimum DBP content resulting in a rubbery state of the PVC-DBP system at 25 °C was estimated to be approximately 30 wt%. Therefore, model material consisting of 30 wt% of DBP in PVC was selected for the MD simulations to fill the gap in the available literature data and evaluate the accuracy of this approach for the prediction of $D$.

MD simulations were performed for the temperature of 25 °C in multiple runs resulting in a continuous trajectory lasting 120 ns. According to the literature, the non-linearity of the MSD in time was observed (Figure 3A), for which three distinct regimes were distinguished: short-time diffusivity (blue linear function), anomalous (orange linear function), and narrow range of the long-term motion (green linear function). The fitted functions for in-cavity and anomalous diffusion exhibited slopes significantly lower than 1, indicating that these processes deviate from the Einstein relation. Long-term diffusion, however, follows Einstein's relation, for which $D$ was estimated as $9.2 \cdot 10^{-14}$ $m^2/s$.

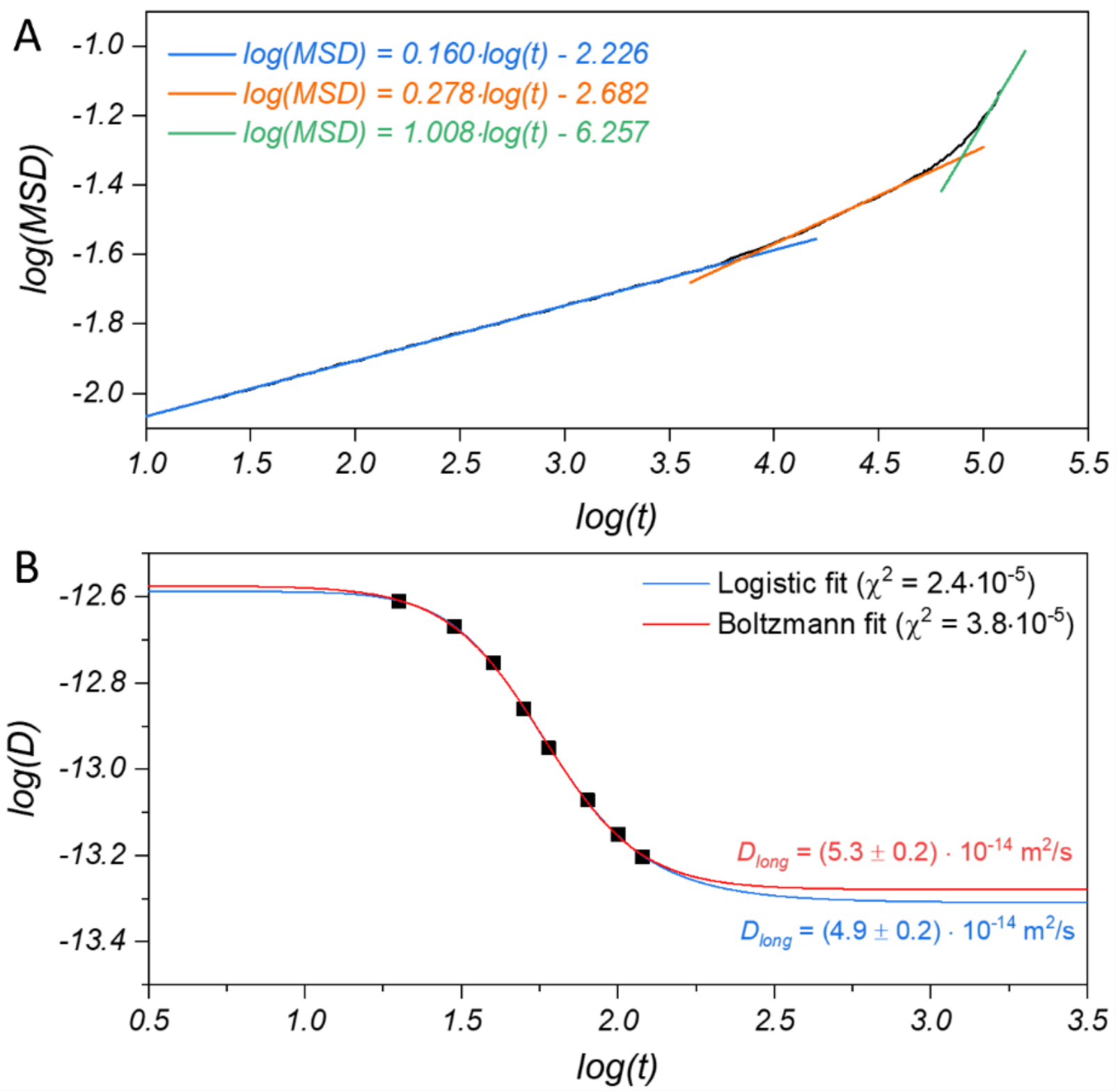

**Figure 3. A**: The evolution of MSD [$nm^2$] in simulation time $t$ [ps] for the DBP (30 wt%) in PVC at 25 °C (298 K); **B**: The dependence of cumulative $D$ [$m^2/s$] values on time [ns] for the DBP (30 wt%) in PVC at 25 °C (298 K).

However, due to the narrow range of normal diffusive motion of DBP plasticiser accessible for MD simulations, an analogous approach for predicting long-term $D_{long}$ following the procedure initially outlined by Blanco and co-workers (Blanco et al., 2018). Assuming isotropic DBP motion in the PVC matrix, cumulative $D$ values were calculated for different periods ranging from 20 to 120 ns (Figure 3B, section A2 in Supplementary Materials). Generally, the diffusivity of small molecules in a polymer matrix is constant in the short- and long-time regimes, whereas in the anomalous range, $D$ is time-dependent. Thus, the relation of $D(t)$ is expected to have a sigmoid shape, where the horizontal asymptotes correspond to the fast in-cavity (upper asymptote) and normal (lower asymptote) diffusion. Bearing this in mind, logistic (blue curve, Figure 3B) and Boltzmann (red curve, Figure 3B) functions were selected and then fitted to the MD data and long-term $D$ values were estimated from the lower asymptotes' values as $4.9 \cdot 10^{-14}$ $m^2/s$ and $5.3 \cdot 10^{-14}$ $m^2/s$, respectively.

To verify this approach and assess the accuracy of the $D$ prediction, solid-state PGSE NMR spectroscopy was applied to measure $D$ of DBP as in the model material, i.e. containing 30 wt% of DBP. The NMR results obtained at 45, 50, 60, 70, 80, 90, 100, and 110 °C (purple rhombus) were compared in the Arrhenius plot (Figure 4) with the predicted MD result from the logistic fit (blue sphere) and data available from the literature (Brouillet and Fugit, 2009; Griffiths et al., 1984).

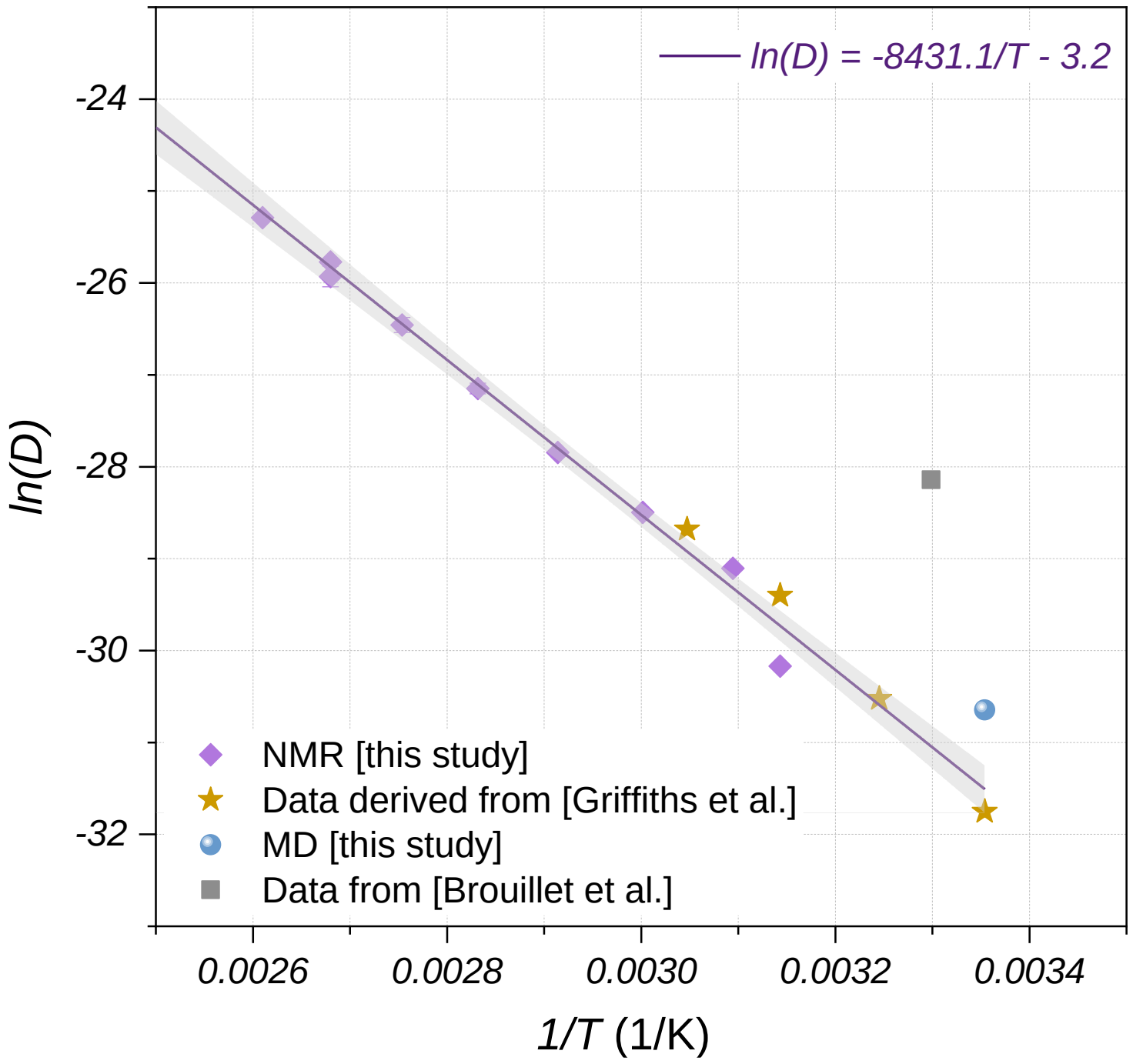

**Figure 4.** Arrhenius plot of DBP diffusion coefficients $D$ [$m^2/s$] for PVC-DBP (30 wt%) system in a rubbery state (> $T_g$). A light grey area along the fitted function indicates a confidence band of 95%.

Firstly, values of $D$ reported by Griffiths and co-workers determined for $^{14}C$-labelled DBP at different weight contents (28.6, 37.5, 44.4, and 50.0 wt%) and various temperatures ranging from 25 °C to 55 °C were used for the recalculation to obtain $D$ values for 30 wt% DBP, as in the model material (for details see section A4 in Supplementary Materials). The resulting values were plotted as beige stars in Figure 4. Generally, $D$ values are considered sufficiently accurate if they are of the same order of magnitude (Gusev et al., 1994). Herein, the NMR results correlate well with the $D$ values estimated from the study by Griffiths and co-workers. Therefore, a linear function was fitted to data obtained from NMR and recalculated from data reported by Griffiths et al. As a result, Equation 2 (dark purple linear fit; units: $D$ [$m^2/s$] and $T$ [K]) allowing estimation of $D$ at 25 °C was obtained:

$$\ln(D) = -\frac{8431.1}{T} - 3.2 \quad (2)$$

Herein, it should be noted that Equation 2 describes the diffusion of DBP solely in a rubbery state (> $T_g$), where the last fitting point measured at 25 °C (beige star at ca. 0.00335 1/K) is considered to be the closest to the $T_g$ and therefore, fitted function was not extended further, where an increase in the slope of *ln(D) = f(T)* is expected corresponding to slower diffusivity manifested by a larger activation energy in the glassy state (see further discussion in section 3.3).

As shown in Figure 4, the $D$ value at 25 °C predicted from the logistic fit to MD results ($4.9\cdot10^{-14}$ $m^2/s$) is in line with the value estimated from Equation 2 ($2.1\cdot10^{-14}$ $m^2/s$), which validates the method of normal (long-term) $D$ prediction based on short-time data via MD simulations. Moreover, the accuracy of this prediction is satisfactory contrary to the $D$ value obtained from the classical extraction method - $6.0\cdot10^{-13}$ $m^2/s$ (grey square) (Brouillet and Fugit, 2009), which indicates an overestimation of $D$ about one order of magnitude. Comparison of sigmoid fitting procedures – logistic and Boltzmann – indicates minor differences in the obtained results. However, $D$ calculated from the lower asymptote of the logistic function is closer to NMR and literature data, thus this procedure is considered more suitable for estimating long-term $D$ values.

The need for reliable values of $D$ from the heritage science as well as industrial perspective is crucial for modelling plasticiser loss and predicting the lifetime of museum objects and commercial products. For this purpose, short-term MD simulations extrapolated with a basic logistic function provide fairly accurate diffusivity values in a reasonable computational time, which is significantly faster than commonly used experimental methods and avoids overestimation of the plasticisers' $D$ in the polymer matrix. Additionally, the application of PGSE NMR spectroscopy opens new paths for characterising heritage objects or mock-up samples. Its main advantage is the simplicity of sample preparation (cutting the sample into the desired shape), chemical non-interference with the investigated material due to lack of solvents, and wide range of operating temperatures (e.g. 10 - 100 $^{o}$C), but the drawback is a destructive character which hinders its wider application in the heritage sector. However, the development of portable NMR instruments (e.g. NMR-MOUSE® (Baias, 2017)) that can be used *in situ* opens a potential path for statistical evaluation of the rate of plasticiser migration in PVC collections.

### 3.3. Rate-limiting process of ortho-phthalate plasticiser migration

To get an insight into the risk of heritage PVC objects related to the plasticiser migration and the potential consequences of conservation treatment on the rate of further deterioration, air pollution, and high-impact mechanical damage to objects becoming stiffer uniformly (evaporation-controlled mode) or only on the outer layer (diffusion-controlled mode), establishing the rate-limiting process of the overall plasticiser loss is a decisive step. Due to a lack of data that enables a comparison of the rate of diffusion and evaporation for all plasticisers present in heritage objects, the simplified approach is adopted based on a comparison of activation energy $E_a$ values of single processes. This simplified approach is a reasonable approximation as the thickness of heritage objects does not vary significantly, typically ranging between 0.2 mm and 10 mm. Briefly, the rate-limiting process (slower one) is defined by a significantly higher $E_a$ value, i.e. plasticiser migration is a diffusion-controlled process, when $E_a$ of diffusion ($E_{a,diff}$) is larger than $E_a$ of evaporation ($E_{a,vap}$) and vice versa. Therefore, values of the $E_a$ of diffusion of plasticiser molecules in the PVC matrix and their subsequent evaporation from the surface were taken into consideration.

Firstly, two diffusion regimes were distinguished depending on the state of the material at given conditions, namely for highly plasticised PVC objects in the rubbery state (above $T_g$) and for slightly plasticised PVC objects in the glassy state (below $T_g$). Values of the $E_a$ corresponding to the diffusion in both states were collected and depicted in Figure 5A for various ortho-phthalate plasticisers as the dependence on their molar mass $M$.

According to Equation 2 obtained in this study for the PVC-DBP system in the rubbery state, the $E_a$ value of DBP diffusion was estimated as 70.1 kJ/mol (purple rhombus, Figure 5A). The $E_a$ values for DBP, DHP and didecyl phthalate – DDP (beige stars, Figure 5A) diffusion in rubbery state (at plasticiser content of ca. 30 wt% ensuring rubbery state) reported by Griffiths and co-workers were determined experimentally and calculated theoretically, respectively: 80.0 kJ/mol and 77.0 kJ/mol for DBP, 85.0 kJ/mol and 77.0 kJ/mol for DHP and 74.0 kJ/mol and 76.0 kJ/mol for DDP (Griffiths et al., 1984). The resulting linear function fitted to these values (dash line, Figure 5A) correlates the $E_a$ [kJ/mol] of diffusion in the rubbery state with $M$ [g/mol] for ortho-phthalates. As shown in Figure 5A, in the case of PVC objects in the rubbery state, the $E_a$ values vary slightly with $M$ and can be assumed as being around 77 kJ/mol for all ortho-phthalate plasticisers since the slope of the fitted function is not statistically different from 0. According to the correlation of $T_g$ with plasticiser content (Figure 2), a minimal concentration of plasticisers resulting in a rubbery-like PVC state at room conditions is around 20-30 wt% depending on the plasticiser type. At higher concentrations of plasticisers, the $E_{a,diff}$ decreases even down to 50 kJ/mol, which leads to a larger difference between the rate of their diffusion and evaporation making diffusion negligible from the overall plasticiser migration perspective. Therefore, the $E_{a,diff}$ of ca. 77 kJ/mol represents the slowest possible diffusion in the rubbery PVC objects.

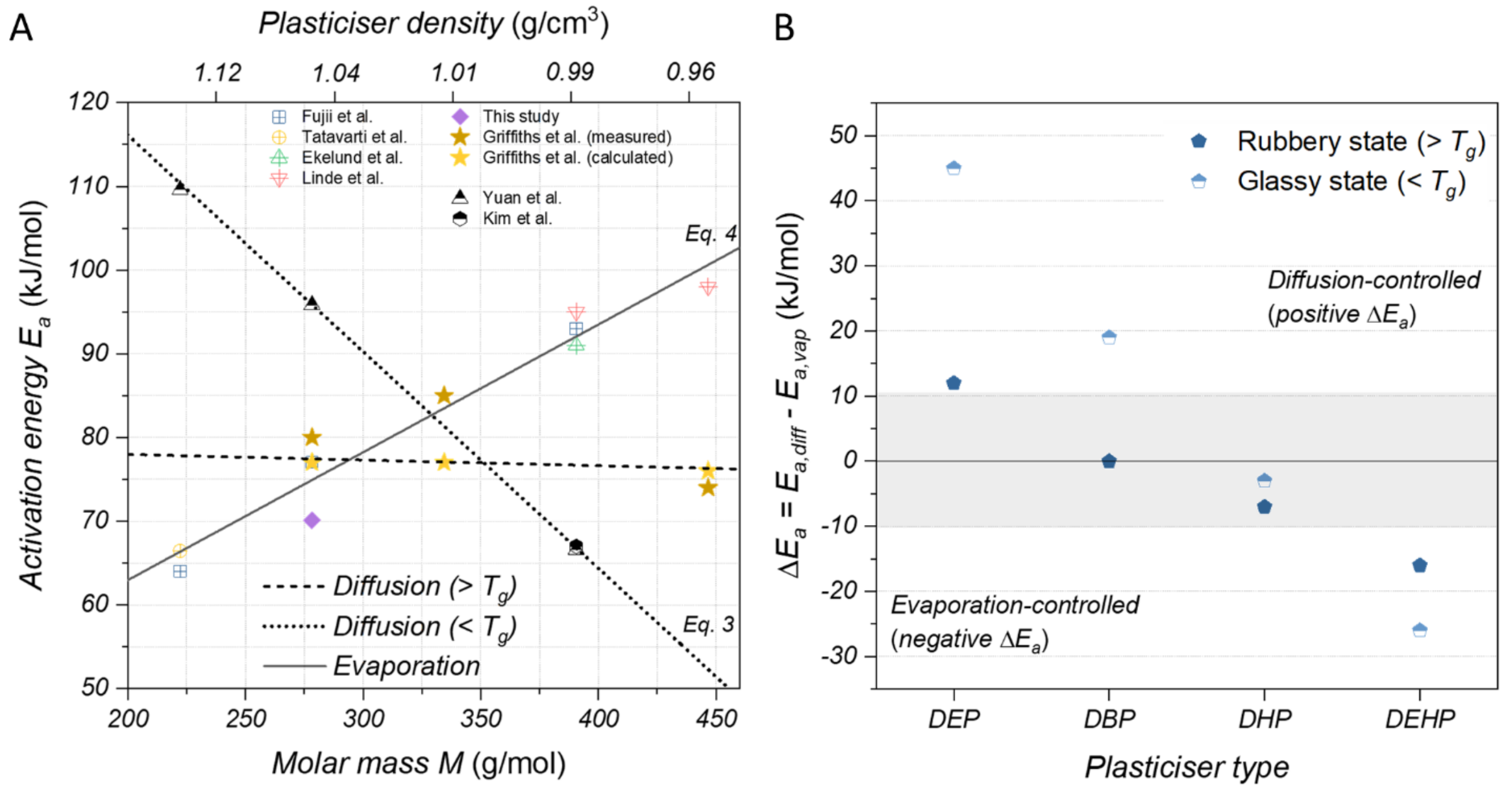

**Figure 5. A:** Correlation between the molar mass $M$ of ortho-phthalate plasticisers (bottom x-axis), their density (top x-axis) and activation energy $E_a$ of their evaporation (solid line, blank-crossed symbols), diffusion in a rubbery (> $T_g$, dash line, full symbols), and diffusion in a glassy (< $T_g$, dot line, half-filled symbols) state. **B:** Difference in activation energy values ($\Delta E_a$) of diffusion ($E_{a,diff}$) and evaporation ($E_{a,vap}$) for the selected ortho-phthalates corresponding to objects in rubbery (dark blue full symbols) and glassy state (light blue half-filled symbols). A bright grey area indicates an insignificant $\Delta E_a$.

Analogously, average values of the $E_{a,diff}$ for the diffusion of ortho-phthalates in the glassy state PVC objects reported in the literature were plotted against $M$: 109.6 kJ/mol, 95.9 kJ/mol and 66.6 kJ/mol for DEP, DBP, and DEHP, respectively (Yuan et al., 2020) and 67.0 kJ/mol for DEHP (Kim et al., 2022). As a result of a linear fit to this data (half-filled black symbols, Figure 5A), the function (dot line, Figure 5A) described by Equation 3 is proposed:

$$E_{a,diffg} = -0.26M + 167.91 \quad (3)$$

A non-intuitive trend between $M$ and $E_{a,diffg}$ represented by a negative slope suggests that diffusion of smaller molecules is slower than diffusion of larger molecules, contrary to the common conviction that small molecules are more mobile. However, the diffusivity of the molecule strongly depends on the density of the investigated system which is related to the free volume of the polymer material (Gusev et al., 1994). Therefore, more densely packed molecules have lower $D$ values and their effective diffusivity becomes slower. The phenomenon is reflected in this particular case, where larger ortho-phthalate molecules exhibit lower density due to the steric effects of longer aliphatic chains (please see top x-axis in Figure 5A), which leads to larger free volume and thus higher effective diffusivity in comparison to smaller ortho-phthalates.

It also suggests that larger molecules, e.g. DEHP (391 g/mol), are more efficient plasticisers than the smaller molecules, e.g. DBP (ca. 278 g/mol). The plasticising efficiency can be assessed as the amount of plasticiser required to reach a certain $T_g$ value. As shown in Figure 2, the function describing $T_g$ dependence on the DEHP content is below the $T_g$ dependence on the DBP content, i.e. to lower $T_g$ of PVC to e.g. 280 K (7 ºC), roughly 30 wt% of DEHP or 40 wt% of DBP is required, what is a consequence of plasticiser structure and steric effects of aliphatic chains reflected in the density of single plasticisers (Figure 5A: the larger the $M$, herein the longer the aliphatic chain substituents, the more efficient as plasticiser ortho-phthalate is).

In the case of highly plasticised PVC resulting in a rubbery state of the material (Figure 5A: dash line), the effect of plasticiser structure on its diffusivity is less distinct and depends rather on the plasticiser content (Griffiths et al., 1984) contributing to the increase in the mean free volume as shown by Borek et al. (Borek and Osoba W., 1996) for DBP in PVC: the mean size of free volume cavities increases slightly up to 25 wt% when the material is still in a glassy state and then changes rapidly upon further DBP addition after the transition to a rubbery state.

Secondly, $E_{a,vap}$ values of ortho-phthalates evaporation (surface emission) were collected from the previous publications (blank-crossed symbols, Figure 5A; exact values are listed in section A5 in Supplementary Materials) (Ekelund et al., 2010; Fujii et al., 2003; Linde and Gedde, 2014; Tatavarti et al., 2002) and a linear concatenated fitting procedure was applied to correlate $E_{a,vap}$ with $M$ (Equation 4):

$$E_{a,vap} = 0.15M + 32.42 \quad (4)$$

Based on the functions shown in Figure 5A, a simple estimation of the rate-limiting step of the overall plasticiser migration can be done for the entire range of ortho-phthalate plasticisers found in heritage PVC collections. Generally, the crosspoint of functions describing diffusion and evaporation in Figure 5A can be considered as a critical point defined by $M_{critical}$, below which plasticiser loss is a diffusion-controlled process ($E_{a,diff}$ >> $E_{a,vap}$) and above which it is an evaporation-controlled process ($E_{a,diff}$ << $E_{a,vap}$). Considering highly plasticised PVC objects, $M_{critical}$ is around 290 g/mol, whereas for slightly plasticised PVC objects the value reaches ca. 330 g/mol, meaning that in the case of ortho-phthalate plasticisers that have larger $M$ than $M_{critical}$, their migration is evaporation-controlled.

However, the main limitation of this concept is that the geometry and dimensions (thickness) of the objects are not taken into account. Therefore, the presented approach enables more likely the identification of whether objects are in diffusion- or evaporation-controlled mode solely when the difference between $E_a$ values of these processes is significant. For this purpose, the difference between $E_{a,diff}$ and $E_{a,vap}$ for selected ortho-phthalates corresponding to artefacts in the rubbery and glassy state was calculated (Figure 5B). Thus, positive and negative $\Delta E_a$ values correspond to diffusion- and evaporation-controlled modes, respectively. Bearing in mind the uncertainty of the $E_a$ determination, the mean uncertainty was estimated as 5.2 kJ/mol based on previously reported literature (Griffiths et al., 1984; Kim et al., 2022; Yuan et al., 2020) (for details please see section A6 in Supplementary Materials). According to this, an insignificant $\Delta E_a$ value was assumed to be double the average uncertainty that was depicted as the light grey rectangle in Figure 5B. Therefore, any $\Delta E_a$ larger than 10.4 kJ/mol and smaller than -10.4 kJ/mol could be considered significant. This simplified approach enables us to appoint PVC-based systems that can be most probably ascribed to the diffusion-controlled migration mode: glassy PVC objects containing DEP ($\Delta E_a$ ca. 45 kJ/mol) or DBP ($\Delta E_a$ ca. 20 kJ/mol). On the other hand, PVC artefacts plasticised with DEHP can be more likely classified as being in the evaporation-controlled mode regardless of DEHP content ($\Delta E_a$ ca. -16 kJ/mol and -26 kJ/mol for objects in rubbery and glassy state, respectively).

Nevertheless, unequivocal determination of the rate-limiting step for certain artefacts requires the determination of surface emission coefficient $F$ [m/s], which would enable precise calculation of critical object thickness based on relation $R=FL/D$ (Wei et al., 2019). When $R >> 1$, plasticiser migration is considered diffusion-controlled and critical thickness $L$ is defined as larger than $D/F$ meaning that if e.g. $D/F$ equals 2 cm, objects thicker than this would be ascribed to the diffusion-controlled plasticiser migration mode.

### 3.4. Practical hints for conservators of plastic artefacts

Based on the above considerations for highly plasticised PVC artefacts (Figure 5B: rubbery state), objects containing higher ortho-phthalates such as DEHP or DINP are most likely in the evaporation-controlled mode. Therefore, they could be potentially cleaned using mild (non-solvent) techniques without increasing the risk of high-impact mechanical damage and presumably diminishing the air pollution in the conservation studio and museum environment by avoiding the uncontrolled release of plasticiser. In the case of 'smaller' plasticisers, i.e. DEP, DBP, and DHP, their diffusion and evaporation are of similar rate (small, insignificant $\Delta E_a$), thus determining the rate-limiting step of their migration will strongly depend on the dimensions of the particular artefacts.

When considering slightly plasticised PVC objects (Figure 5B: glassy state) that are the most representative of heritage collections (mean plasticiser distribution is around 18 wt% which corresponds to the $T_g$ of ca. 50 °C) (Rijavec et al., 2022), DEP-, DBP- or DiBP-containing artworks might be at higher risk of high-impact mechanical damage (possible non-uniform stress distribution) due to their migration being controlled rather by diffusion, which is expected to accelerate upon removal of the thin surface plasticiser layer (usually referred to as rather tacky surface than visual liquid deposits). However, it should be noted that these plasticisers constitute less than 10% of the PVC in heritage collections.

According to the survey reported by Rijavec and co-workers (Rijavec et al., 2022), statistically, 90% of the PVC artefacts contain larger ortho-phthalates (DEHP, DOTP or DINP) and thus can be most likely ascribed to the evaporation-controlled plasticiser migration mode (regardless of their content), which:

- suggests the safety of the mild cleaning techniques (e.g. dry swabbing) that most probably do not affect the migration rate and would diminish dust deposition;
- provides meaningful insight into the risk of mechanical damage to PVC objects which states that the vast majority of them are expected to be at low risk of high impact-induced cracking, e.g. during transportation;
- may contribute to a decrease in air pollution in the indoor environment of heritage institutions by controlled removal of the surface plasticiser layer, which is in line with the preservation strategy.

## 4. Conclusions

This paper discusses the ortho-phthalate plasticisers' migration in the heritage PVC objects according to the general diffusion-evaporation model and attempts to provide evidence-based recommendations for museum practitioners regarding mild cleaning and its consequences on mechanical and environmental aspects. When facing practical conservation issues and decision-making processes concerning collection care, the main requirement is to use fast, simple and accessible tools for museum practitioners. In particular, the simple step-by-step protocol supporting conservators of modern and contemporary art dealing with the deterioration of PVC objects was proposed as follows:

(I) spectroscopic determination of <u>plasticiser type and content</u> using a handheld FTIR or Raman spectrometer (commonly available in museums and heritage institutions) with the support of machine learning-assisted analysis described previously (Rijavec et al., 2022; Saad et al., 2024);

(II) estimation of the state of heritage PVC object, i.e. rubbery or glassy, depending on plasticiser concentration $C$ (and type) based on function $T_g = f(C)$ proposed in section 3.1 of this paper (specifically Figure 2);

(III) determining the rate-limiting step of overall plasticiser migration based on correlations proposed for various ortho-phthalate plasticisers as the most common in heritage collections (described in section 3.3 of this paper; specifically Figure 5B).

Based on this, it was assessed that the majority of heritage PVC objects (ca. 90%) are expected to be at low risk of high-impact mechanical damage and removal of surface plasticisers' deposits with mild cleaning techniques would not affect the further migration rate. Therefore, this approach suggests that most objects could be safely cleaned diminishing dust deposition and contributing to a decrease in air pollution. The latter is particularly important from the view of entire collection care due to the reduction in potential deposition of plasticisers' vapours and/or their degradation products on the other artworks displayed or stored in the vicinity of phthalate-emitting objects.

Additionally, this paper provides a critical evaluation of the use of MD simulations and PGSE NMR spectroscopy as faster alternatives for the determination of plasticisers' diffusion coefficients in PVC matrix, which meets the need for reliable $D$ values that are crucial for modelling plasticiser loss. The procedure based on simple logistic fitting to the short-time MD data for predicting long-term $D$ values was proposed and assessed as accurate providing the $D$ values within the same order of magnitude according to the NMR study and literature data. Therefore, instead of long-lasting extraction or accelerated ageing methods, MD simulations can be successfully applied as a fast and non-destructive approach avoiding overestimation of $D$ values. Together with the determination of surface emission coefficients $F$ specific for each ortho-phthalate, it would provide a comprehensive overview of the plasticiser migration in PVC heritage collections, which is a subject of our further research.

**Supplementary Materials**

Details on the dynamic mechanical-thermal analysis (DMTA) (section A1), MD simulations (section A2), pulsed-field gradient spin-echo nuclear magnetic resonance (PGSE NMR) spectroscopy (section A3), recalculation of $D$ for 30 wt% of DBP based on literature data (section A4), activation energy of plasticisers evaporation (section A5), and evaluation of the typical $E_a$ uncertainty (section A6) are available in the Supplementary Materials.

**Acknowledgements**

The studies were financially supported by the PVCare project *Preventive Conservation Strategies for Poly(vinyl chloride) Objects* funded through the OPUS-LAP 20 programme by NCN (National Science Center, Poland, project no. 2020/39/I/HS2/00911) and ARIS (Slovenian Research and Innovation Agency, project no. N1-0241).

The authors are grateful to the Cricoteka - Tadeusz Kantor Art Documentation Centre in Cracow (Poland) for providing access to the collection and allowing high-resolution scanning of the selected objects. The authors also thank Anwil SA (Orlen Group, Poland) for providing a sample of powdered PVC.

**Declaration of Competing Interests**

The authors declare that they have no competing interests.

**CRediT author statement**

**Sonia Bujok:** Conceptualization, Formal analysis, Investigation, Resources, Data Curation, Visualization, Writing - Original Draft, Writing - Review & Editing
**Tomasz Pańczyk:** Methodology, Formal analysis, Investigation, Data Curation, Visualization, Writing - Original Draft
**Kosma Szutkowski:** Methodology, Formal analysis, Data Curation, Writing - Original Draft
**Dominika Anioł:** Formal analysis, Investigation, Data Curation, Visualization, Writing - Original Draft
**Sergii Antropov:** Investigation, Visualization
**Krzysztof Kruczała:** Writing - Review & Editing, Project administration, Funding acquisition
**Łukasz Bratasz:** Investigation, Writing - Review & Editing, Project administration, Funding acquisition

**Supplementary Materials**

**A1. Dynamic mechanical thermal analysis (DMTA)**

Figure S1 shows the temperature dependence of the storage modulus *E'* (tensional) and *tanδ* of the unplasticised PVC (A) and PVC plasticised with 20 and 30 wt% of DBP (B and C, respectively).

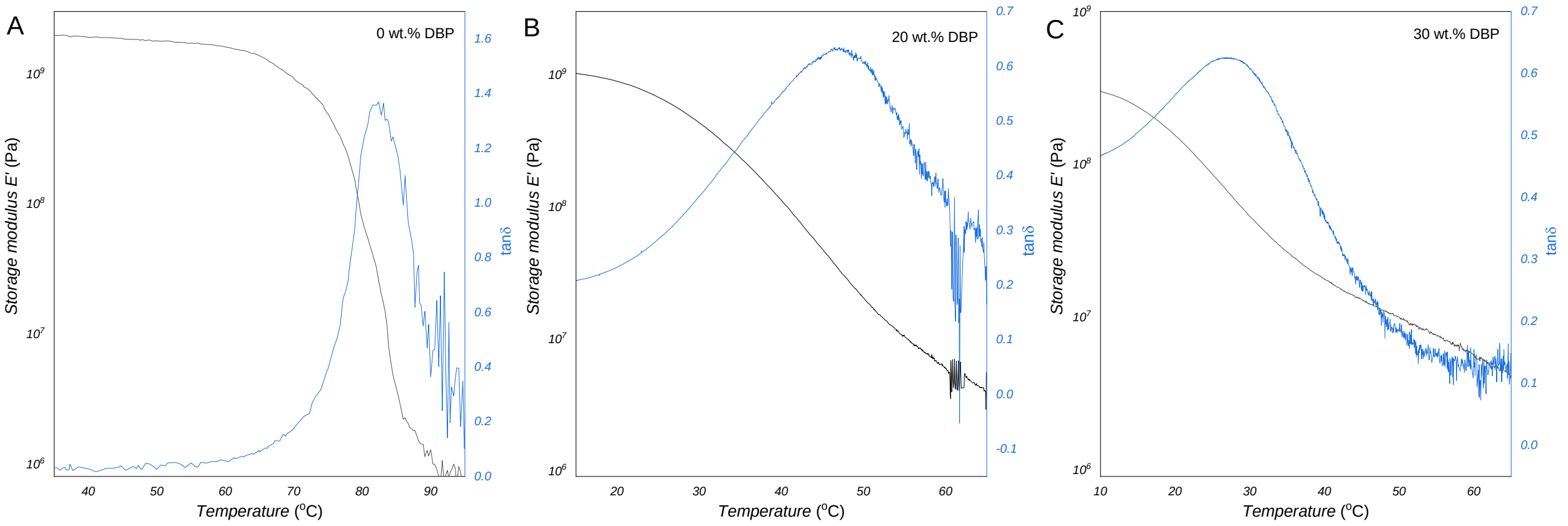

**Figure S1.** Temperature scans of unplasticised (A) and plasticised PVC samples (B: 20 wt%, and C: 30 wt%).

**A2. Construction of the simulation box for molecular dynamics simulations**

The prototype of the PVC molecule (Figure S2) composed of 7 repeating units was built manually and used to calculate partial atomic charges. The charges were calculated using quantum chemical determination of electrostatic potential on the Hartree-Fock level of theory and further fitting of this potential to atomic charges using the RESP-A1 procedure (Dupradeau et al., 2010; Vanquelef et al., 2011). The resulting geometry and charges of the prototype PVC subchain are the following:

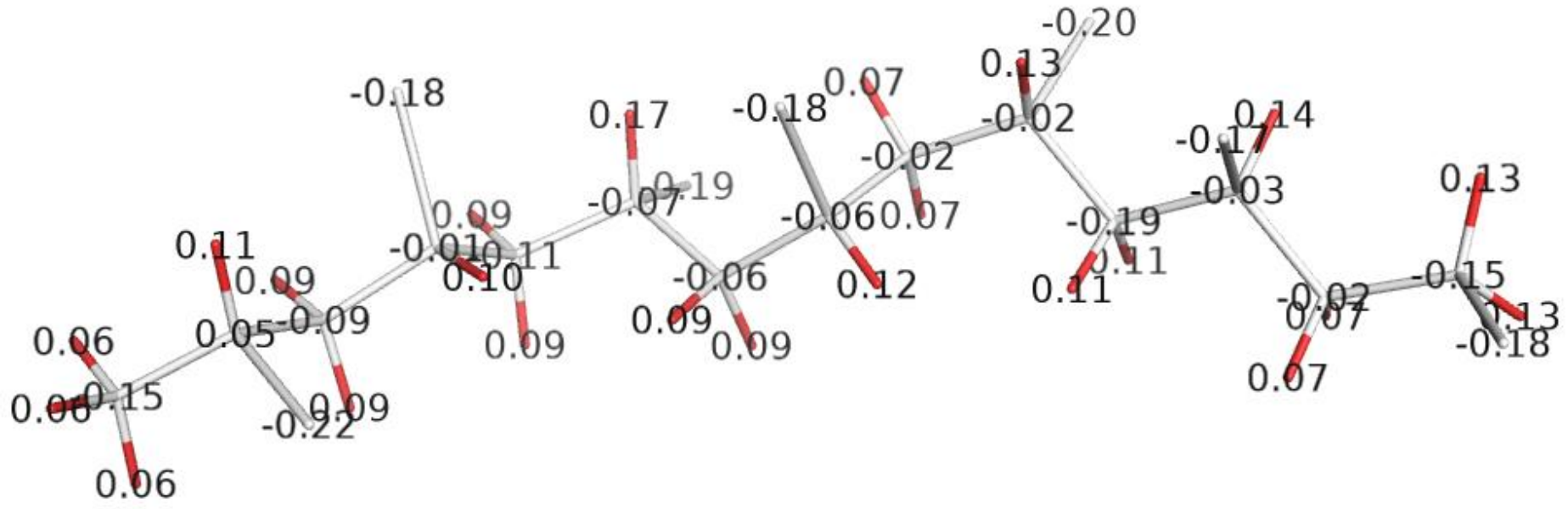

**Figure S2.** The prototype of a PVC subchain consisting of a sequence of 7 repeating units ($CH_2$-CHCl).

The terminal segments of the PVC prototype were removed and the resulting sequence of 5 repeating units was used for building a PVC macromolecule by repeating this unit 1490 times and adding the end groups of the PVC that were capped by previously removed terminal segments. Thus, the resulting single macromolecule of PVC has all saturated bonds and zero total charge. The molar mass of a single PVC macromolecule is 93243 g/mol, which is comparable to the molar mass of PVC representative in heritage collections (Rijavec et al., 2023). The force field for MD calculations was assumed to be the most recent Amber (parm10) (Case et al., 2023), which is commonly used in simulations of new organic molecules (a typical pairwise additive force field). The molecular structure of dibutyl phthalate (DBP) was also constructed manually and next subjected to quantum chemical calculations to determine the geometry and partial charges (Figure S3), analogically to the prototype PVC subchain.

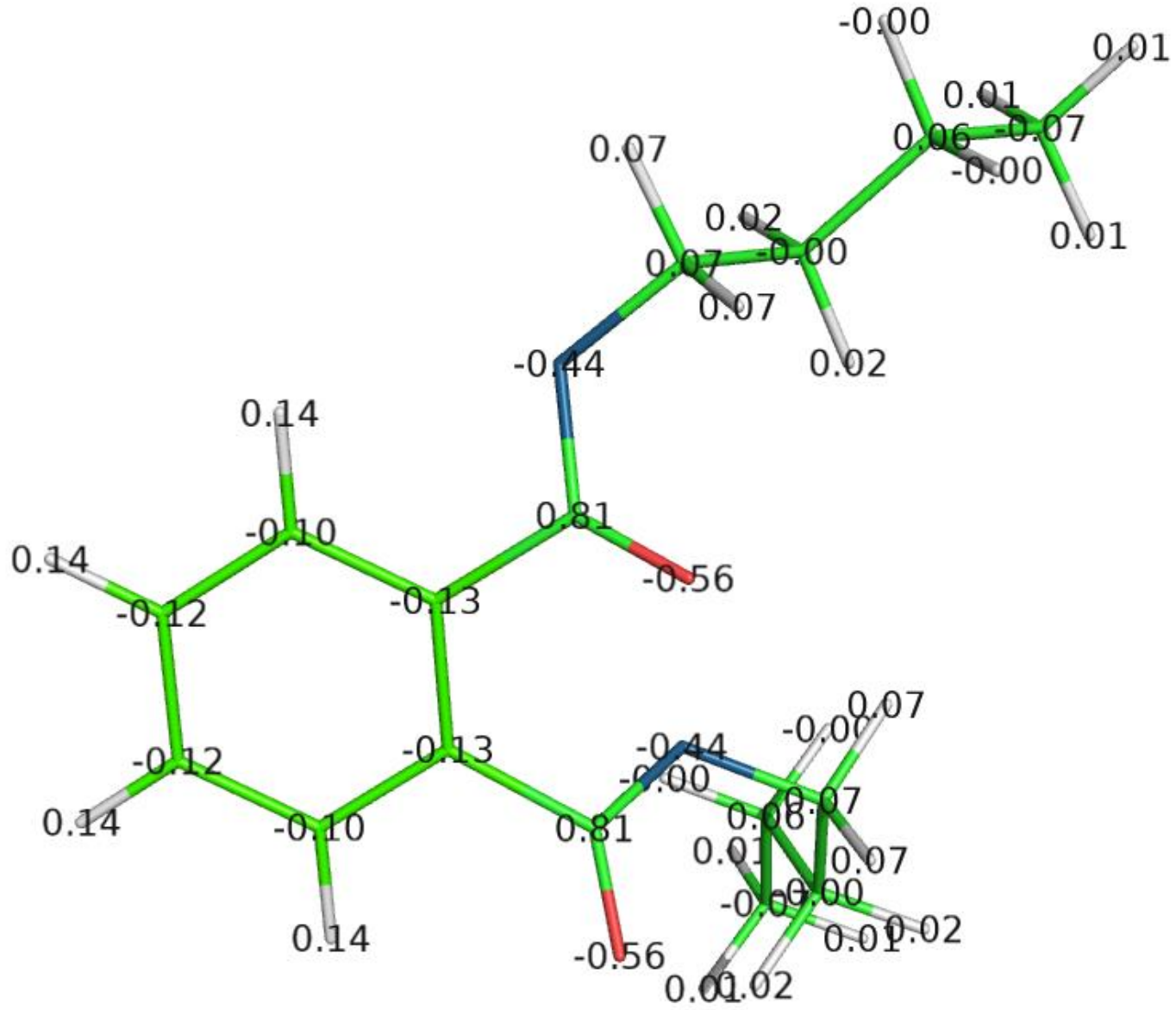

**Figure S3.** Structure and charges of the DBP molecule.

The simulation box was constructed using 4 PVC macromolecules and 572 DBF molecules resulting in the DBP weight content of ca. 30% and the total number of atoms in the simulation box of 59840. At the initial state, each of the PVC macromolecules was a straight linear chain aligned parallelly to each other. The DBF molecules were inserted at equal distances from each other and formed also a kind of linear chains oriented almost parallel to PVC chains. Thus, the initial geometry of the simulation box was: four parallel chains of PVC and four stacks of DBP molecules aligned one by one.

The initial configuration of the system underwent compression, specifically, a simulation in the NPT ensemble, to achieve an adequate density of the system at p = 1 bar and T = 298 K. Subsequently, thermal annealing was performed to eliminate any local energetic barriers formed during the compression. Finally, the system was brought back to 298 K in the NPT ensemble and allowed to relax for 20 ns. The resulting configuration of the system is depicted in Figure S4. As observed, it exhibits a completely amorphous and atactic structure because no additional constraints were imposed to produce locally crystalline and syndiotactic areas. This is consistent with the general understanding that plasticised PVC is amorphous, although some degree of crystallinity has been observed in both rigid and plasticised PVC (Gilbert, 1994). The determined density of the model system at 25 ºC (1.296 g/cm$^3$) is in line with the density of PVC containing 30 wt% of DBP reported in the literature (1.291 g/cm$^3$) (Dlubek et al., 2003), which indicates that the material structure of the model system is well described. Therefore, this system was used to study the diffusivity of DBP molecules (red spheres) trapped in vacancies created by PVC macromolecules (green spheres).

The full production run spanned a trajectory of 120 ns and comprised six 20 ns runs, each initiated from the final configuration of the preceding run. Diffusivities were determined using the GROMACS tool gmx_msd, which computes diffusivity from the Einstein relation using built-in settings. Table S1 displays the diffusivities determined from the full 120 ns trajectory at various time endpoints to demonstrate that diffusivity varies over time. Simulation settings are summarized in Table S2.

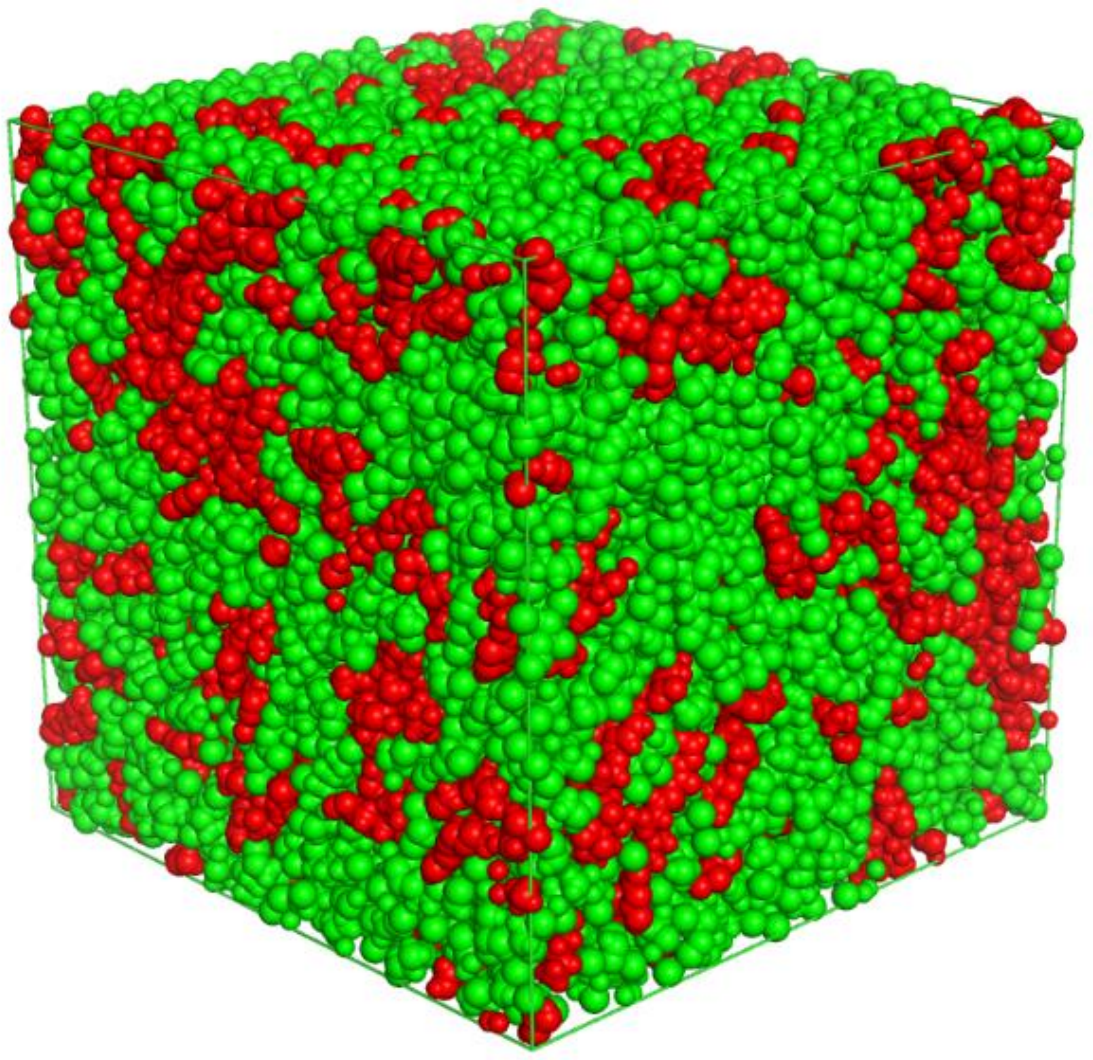

**Figure S4.** Visualisation of the equilibrated simulation box, where red and green spheres refer to DBP and PVC molecules, respectively. The visualised configuration corresponds to 25 °C (298 K) and 1 bar and is the starting point for the diffusivity determination.

**Table S1.** Diffusivity *D* of DBP in PVC at 25 °C (298 K).

| Run no. | Total run time (ns) | Cumulative DBP diffusivity *D* ($m^2/s$) | *log(t)* | *log(D)* |
|---|---|---|---|---|
| 1 | 20 | $2.449 \cdot 10^{-13}$ | 1.301 | -12.611 |
| 2 | 30 | $2.144 \cdot 10^{-13}$ | 1.477 | -12.669 |
| 3 | 40 | $1.763 \cdot 10^{-13}$ | 1.602 | -12.754 |
| 4 | 50 | $1.378 \cdot 10^{-13}$ | 1.699 | -12.860 |
| 5 | 60 | $1.12 \cdot 10^{-13}$ | 1.778 | -12.951 |
| 6 | 80 | $8.48 \cdot 10^{-14}$ | 1.903 | -13.072 |
| 7 | 100 | $7.06 \cdot 10^{-14}$ | 2.000 | -13.151 |
| 8 | 120 | $6.26 \cdot 10^{-14}$ | 2.079 | -13.203 |

**Table S2.** Simulation settings.

| | |
|---|---|
| Box dimension after compression | 8.833x8.833x8.833 $nm^3$ |
| Total number of atoms | 59840 |
| Thermostat | V-rescale |
| Thermostat time constant | 0.1ps |
| Barostat | Parrinello-Rahman |
| Barostat time constant | 2.0ps |
| Compressibility | $4.5\ 10^{-5}\ bar^{-1}$ |
| Integration timestep | 0.001fs |
| Electrostatics | particle-mesh-Ewald |
| Periodic boundary conditions | xyz |
| LJ and Coulomb cutoff | 1.2nm |

## A3. Pulsed-field gradient spin-echo (PGSE) NMR spectroscopy

The purpose of this study was to determine the diffusion coefficient *D* of DBP in a model PVC sample containing 30 wt% of DBP using PGSE NMR diffusometry with the Dbppste (DOSY bipolar pulse pair stimulated echo) pulse sequence (Figure S5). The data were initially analysed using MestReNova 14.3. The spectra were then phased and the baseline was corrected (Figure S6A). Subsequently, the signal amplitudes were analysed using OriginPro 2024 where a linear fit was applied to the logarithm of the signal intensities *M(g)* versus squared magnetic field gradient ($g^2$). Accordingly, the diffusion coefficients *D* of the DBP were derived from the slope using the linearised Stejskal-Tanner equation (Figure S6B) (Stejskal and Tanner, 1965):

$$ln \frac{M(g)}{M(g=0)} = -D(\gamma^2\delta^2(\Delta - \delta/3))\, g^2 \quad (eq.S1)$$

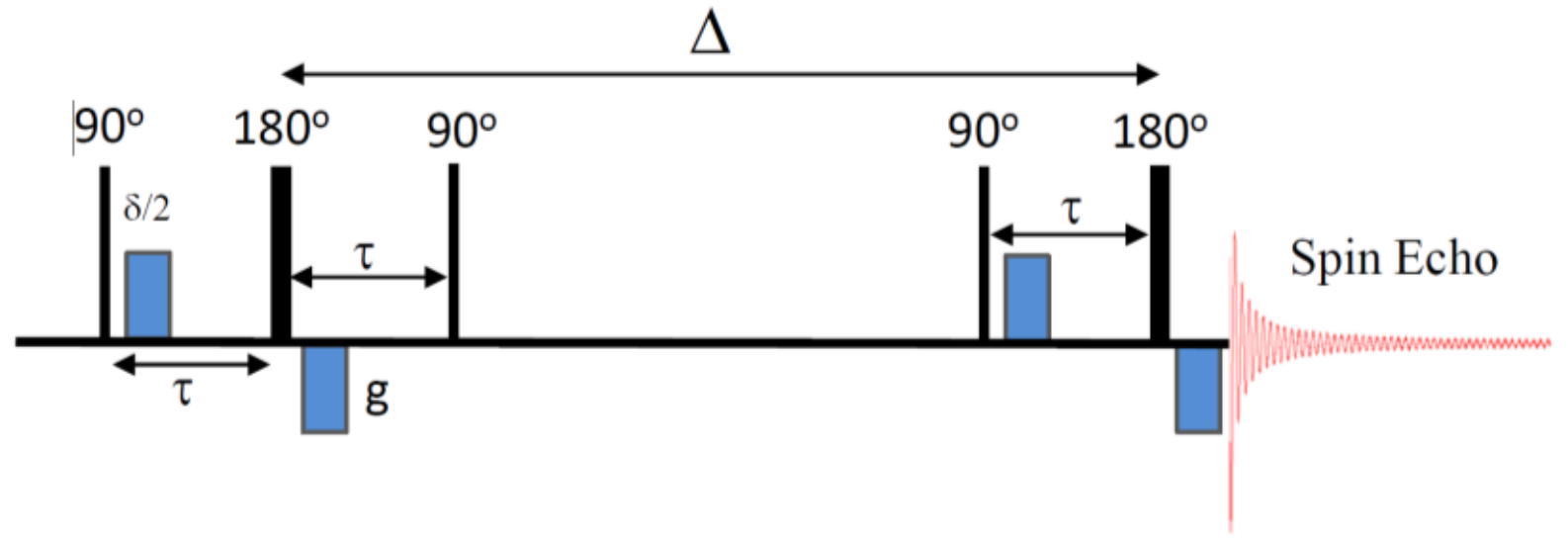

**Figure S5.** Dbppste pulse sequence.

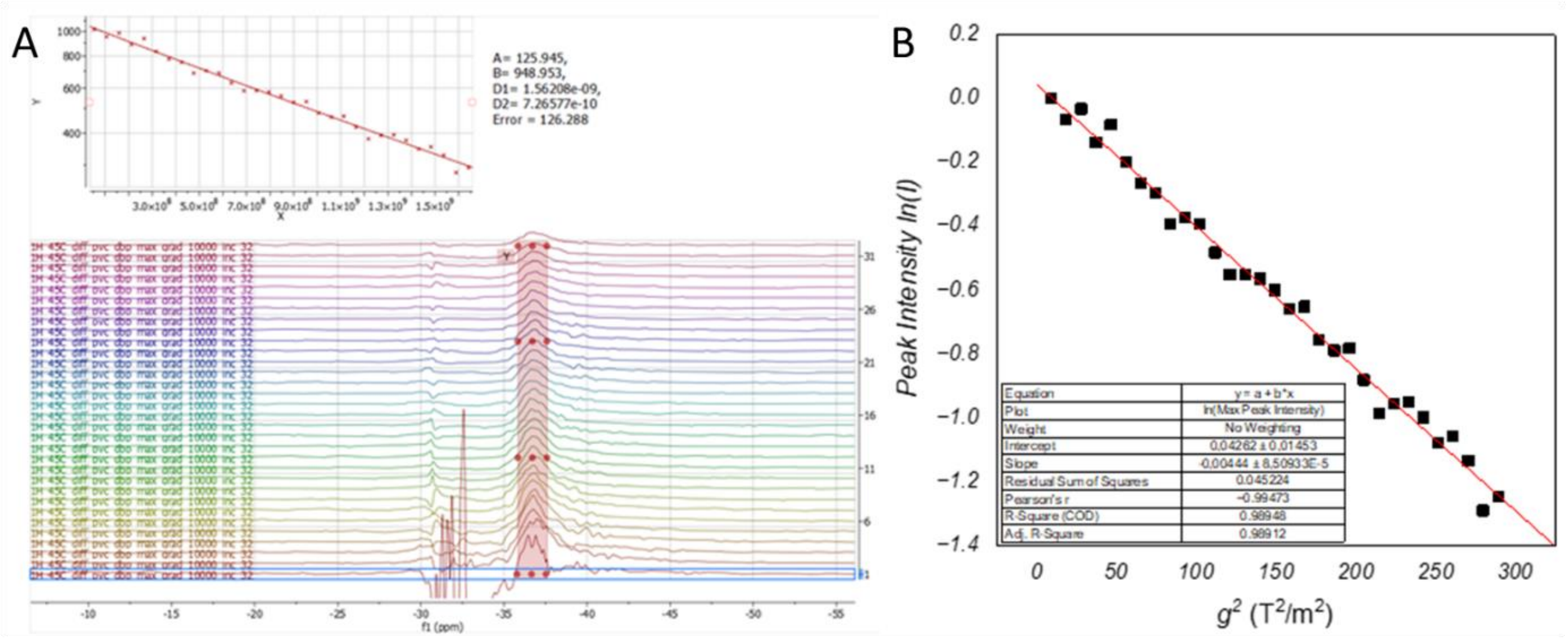

**Figure S6.** Exemplary NMR spectra (A) derived from Dbppste experiment and function fit (B) to the data collected at 45 ºC.

## A4. Recalculation of *D* at 25, 35, 45, and 55 °C for 30 wt% of DBP based on literature data

Data shown in Figure 4 as beige starts were estimated on the basis of linear fitting to the experimental data reported by Griffiths and co-workers (Griffiths et al., 1984) as presented in Figure S7.

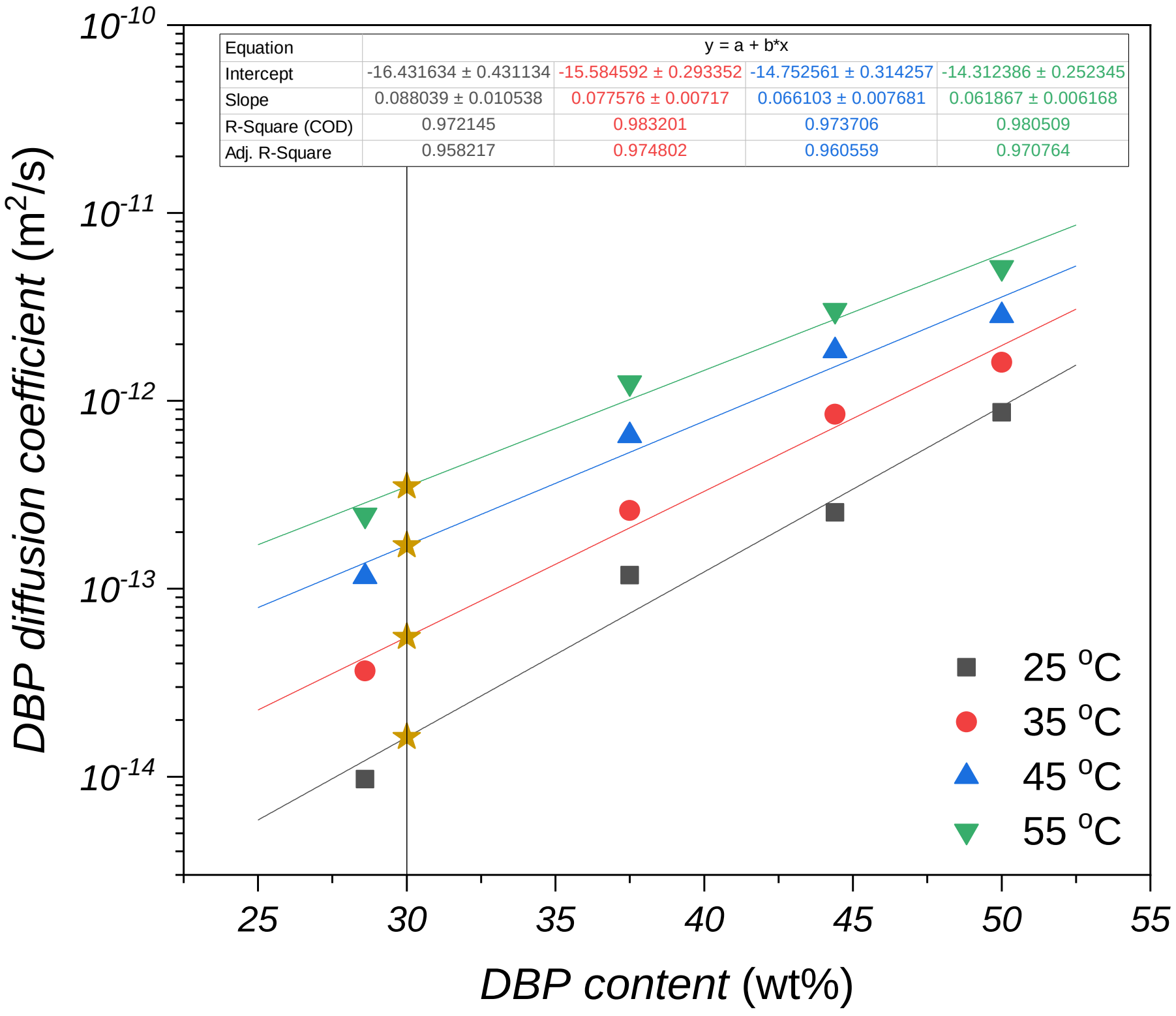

**Figure S7.** Experimental *D* values reported elsewhere (Griffiths et al., 1984) for DBP in PVC at different DBP content (28.6, 37.5, 44.4, and 50.0 wt%) and selected temperatures 25 (black squares), 35 (red circles), 45 (blue upward triangles), and 55 °C (green downward triangles). *D* values estimated for 30 wt% of DBP are marked with beige starts as in Figure 4.

## A5. The activation energy of ortho-phthalates evaporation

Activation energy values of the evaporation of various ortho-phthalate plasticisers previously reported in the literature (Ekelund et al., 2010; Fujii et al., 2003; Linde and Gedde, 2014; Tatavarti et al., 2002) are listed in Table S3.

**Table S3.** Activation energy $E_a$ values of the ortho-phthalates evaporation (surface emission).

| Plasticiser | $E_a$ value [kJ/mol] | Reference and corresponding blank-crossed symbols in Figure 5A |
|---|---|---|
| DEP | 64.0 | Fujii et al. (blue squares) |
| | 66.45 | Tatavarti et al. (yellow circle) |
| DBP | 77.0 | Fujii et al. (blue squares) |
| DEHP | 93.0 | Fujii et al. (blue squares) |
| | 91.0 | Ekelund et al. (green upward triangle) |
| | 95.0 | Linde et al. (red downward triangle) |
| DIDP | 98.0 | Linde et al. (red downward triangle) |

## A6. Evaluation of the typical $E_a$ uncertainty

Because literature data on $E_a$ values are usually reported without providing information on the fit uncertainty which is crucial for discussion on the significance of differences in $E_a$ values, selected available datasets (Griffiths et al., 1984; Kim et al., 2022; Yuan et al., 2020) were analysed (Figure S8) to evaluate the typical $E_a$ uncertainties that are listed in Table S4. Mean $E_a$ uncertainty was calculated as 5.2 kJ/mol.

**Table S4.** Uncertainties of the $E_a$ determination.

| Plot | Dataset | $E_a$ value [kJ/mol] | $E_a$ uncertainty [kJ/mol] | Reference |
|---|---|---|---|---|
| A | 28.6% DHP | 83.6 | 2.5 | Griffiths et al. |
| | 37.5% DHP | 74.7 | 2.0 | |
| | 44.4% DHP | 58.9 | 3.7 | |
| | 50% DHP | 50.2 | 7.1 | |
| B | 28.6% DDP | 74.8 | 5.2 | |
| | 37.5% DDP | 74.5 | 10.8 | |
| | 44.4% DDP | 58.8 | 6.6 | |
| | 50% DDP | 45.5 | 3.0 | |
| C | 1% DEHP_n-hexane | 77.7 | 6.7 | Yuan et al. |
| | 1% DEHP_ethanol | 72.0 | 1.4 | |
| D | 4% DEHP | 66.8 | 2.1 | Kim et al. |
| | 12% DOTP | 97.9 | 11.5 | |
| | | **Mean $E_a$ uncertainty [kJ/mol]** | **5.2** | |

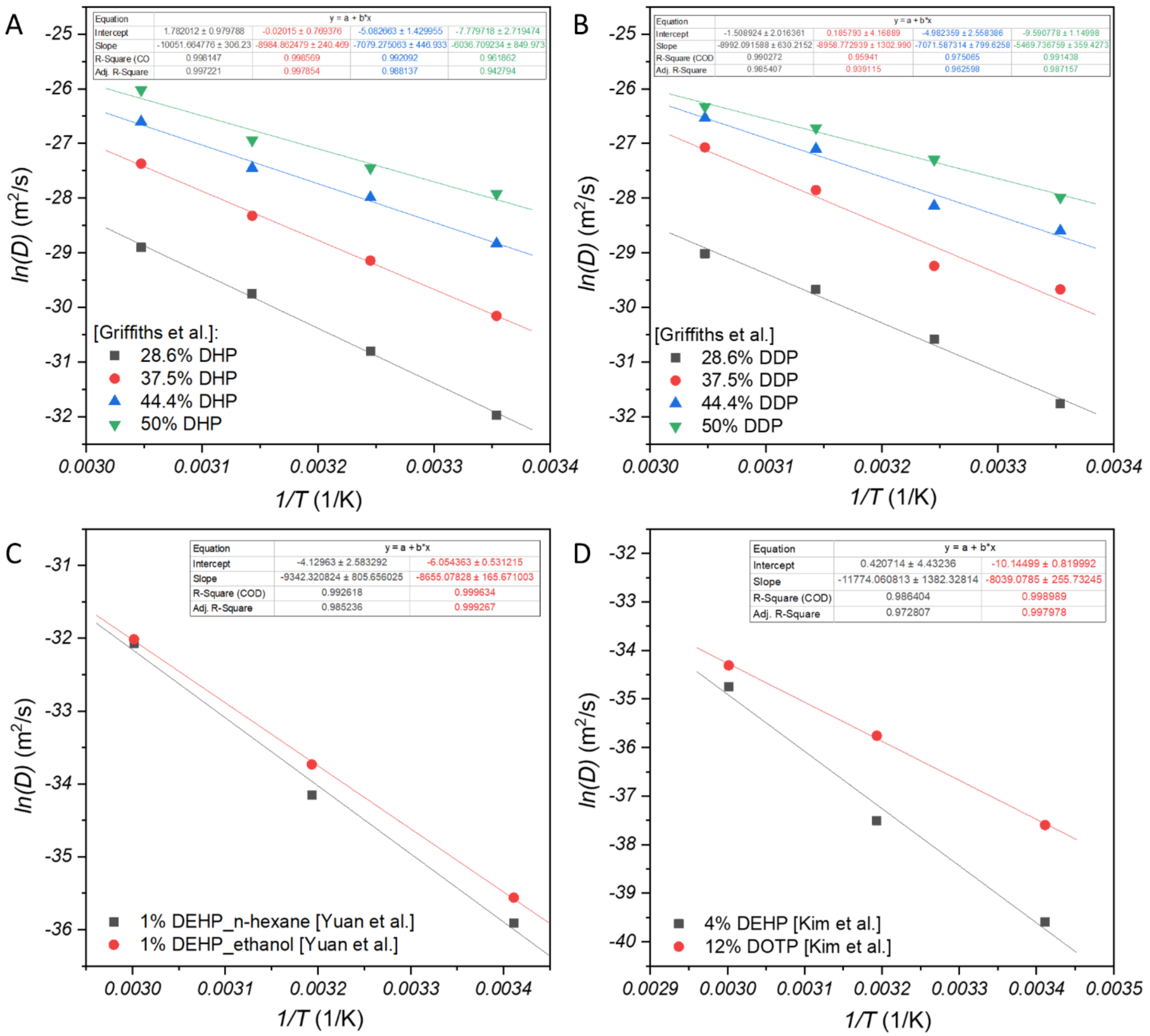

**Figure S8.** Arrhenius plots of phthalates plasticisers' $D$ reported previously by (Griffiths et al., 1984; Kim et al., 2022; Yuan et al., 2020) used for evaluation of $E_a$ uncertainty.